\documentclass{ws-ijmpd}

\newcommand{\ba}{\begin{array}}
\newcommand{\ea}{\end{array}}
\newcommand{\bd}{\begin{displaymath}}
\newcommand{\ed}{\end{displaymath}}
\newcommand{\be}{\begin{equation}}
\newcommand{\ee}{\end{equation}}
\newcommand{\bea}{\begin{eqnarray}}
\newcommand{\eea}{\end{eqnarray}}

\def\q2 {q^2}

\begin{document}


\title{Mixed Phase in Compact Stars : M-R relations and Radial
  oscillations}     
\author{Ashok Goyal \footnote{E-mail: agoyal@iucaa.ernet.in}}
\address{Department of Physics \& Astrophysics, \\
         University of Delhi, \\
         Delhi - 110007, India}

\maketitle

\begin{abstract}
It is believed that quark stars or neutron stars with mixed phase in
the core have smaller radii 
compared to ordinary compact stars. With the recent observation of
several low radius objects, 
typically a radius of $<10 Km.$ for star of mass $< 1M_0$
 in low mass X-ray binaries (LMXB) , it has become
 very important to 
understand the nature of these objects. An accurate determination of
mass-radius relationship of these  
objects provide us with a physical laboratory to study the composition
of high density matter and the nature of phase transition. We study
the effect of quark and nuclear matter  mixed phase on mass radius
relationship and radial oscillations  of neutron stars. We find that the 
effect of the mixed phase is to decrease the maximum mass  of a stable
neutron star and to decrease the radial frequencies .
\end{abstract}

\section{Introduction}
Neutron stars are one of the most fascinating compact objects in the Universe coming into
existence at the end of the evolutionary journey of a massive star. They support 
themselves against gravitational collapse by the pressure of degenerate neutrons and
 are the most
compact and dense stars known. They typically compress a solar
mass matter into a tiny radius of 10 Km with densities in the
core reaching several times the nuclear density. They thus provide a unique opportunity to study
properties of matter, its composition and phases, at extremely  high densities and to perform tests 
of General theory of relativity (GTR). With such densities in
the core, they themselves can take various forms, for example they
could be composed of normal nuclear matter with hyperons and/or
condensed mesons. The matter at such densities may undergo phase
transition to constituent quark matter. It would then be energetically
profitable for the u-d matter to convert itself into u-d-s matter
through weak interactions thereby lowering its energy per baryon. If
it so happens that the energy per baryon of such matter called {\bf
strange matter} (SQM) {\sl i.e.,} matter with roughly equal number of
u, d and s  
quarks with electrons to guarntee charge neutrality, is the true 
ground state of matter with energy per baryon less
than that of iron (939 MeV) the most stable nuclei, the whole star
will convert itself into what is called a {\bf strange star } with
vastly different characteristics. Further, since at large densities 
quantum chromodynamics (QCD) becomes a weakly interacting theory of
quarks and gluons,  
the attractive force between quarks near the Fermi surface due to one
gluon exchange 
 will result in the formation of Cooper pairs giving rise to a {\bf
   Color Superconducting State } 
 wherein the Cooper pairs will condense breaking the color gauge
 symmetry \cite{1}. In general QCD at 
high densities has a very rich phase structure, but the pairing
pattern formed at  
sufficiently high densities for massless quarks is the {\bf color
  flavored locked (CFL)} phase 
in which quarks of all three colors and flavors are paired in a single
condensate. 
In this phase SQM is neutral and no electrons are present. If the mass
of strange quark is not  
too large , this phase may extend to even low densities but for
$m_s\sim150 MeV$, one gets a two 
flavor super conductor (2SC) phase in which only u and d quarks of two
colors are paired to  
form a condensate and the third color pairs with strange quark. 
 In case strange matter is not the
true ground state, the neutron star may have a quark core followed by
a mixed phase and nuclear mantle at the top. Such stars are called
{\bf hybrid stars }.

 A neutron star comes into existence through a
cataclysmic process where all the fundamental forces of nature come
into play. The time scale is large compared to not only the strong and
electromagnetic interactions time scale but weak interaction time
scale as well. Further the electrostatic repulsion being so much
stronger compared to gravitational attraction, the matter is
electrically neutral and typical Fermi momenta of constituents being
large compared to its temperature, neutron star is composed of {\bf
cold, degenerate, charge neutral matter in $\beta$-equilibrium}. At
densities  $< 2 \times 10^{-3}\rho_0$  ($\rho_0=0.16~ nucleons~/fm^3~$ is
the equilibrium density of charged nuclear matter in nuclei), the
matter is assumed to be in the form of a Coulomb lattice (to minimize
energy) of nuclei immersed in a relativistic degenerate electron gas. In the
lower part of the density range $10^{-3}\rho_0~<~\rho~<10\rho_0~$ the
neutrons leak
 out of the nuclei and a Fermi liquid
of neutrons, protons and electrons start building up. At higher
densities there are several possibilities including the occurance of
muons, condensation of negatively charged pions and kaons, appearance
of hyperons and finally, the transition to quark matter.

Matter at such high densities has not been produced in the laboratory
and there is no available data on nuclear matter interactions. The
quantities of interest are the phases and composition of neutron star
matter, its energy density and pressure which determine the equation
of state (EOS). The EOS upto nuclear densities is fairly well
accounted for on the basis of measured nuclear data and nucleonic
interactions. Above nuclear densities there are basically two
approaches: one is to take the interaction between constituents from
realistic fitting with known scattering data and then use the
techniques of many body theory to calculate correlations. The other is
 to take a relativistic mean field type of model with couplings
treated as parameters to fit observable quantities. Both approaches
suffer from lack of experimental data. Whereas many body approach is
well understood, two nucleon interactions are fairly well known,
higher body interactions are not well characterised and the approach
is non-relativistic. Mean field theoretical models can easily
incorporate many constituents, but the complicated correlations are
simplified in terms of vacuum expectation of mean fields which are
fitted to insufficient data.

As discussed above, if there is a phase transition to quark matter,
the entire star may convert itself into a strange star or a hybrid
star depending on whether the strange matter is the true ground state
of matter or not. This has indeed shown to be possible in the MIT bag
 model with realistic values of the parameter wherein the long range 
cofining QCD interactions are taken into account phenomenologically by
 bag pressure and short range interactions perturbatively. Recently a number 
of other models\cite{2} such as the Effective Mass Bag Model , density dependent
quark mass model , a model by Dey et al. in which the SQM has asymptotic
 freedom at high densities  and confinement at zero density built in have been 
considered. In Dey's model, the quark interaction is described by a 
color-Debye-screened interquark vector potential arising from gluon exchange
 and density dependent scalar potential which restores chiral symmetry at 
high densities. In the event
 of strange matter not being the absolute ground state, as happens, for 
example, in the Nambu-Jona-Lasinio model for parameters fitted from low
 energy spectroscopy, or in the MIT model for different values of
the parameters, the quark matter in normal phase or in the color superconducting
 phase as discussed above can exist in the inner regions of 
more massive stars or in the mixed phase in equilibrium with the
confined hadronic phase or both. In earlier studies the phase tramsition
was characterised as a first order transition with a single component
viz baryon number and charge neutrality was strictly enforced
in each phase separately. This gave rise to constant pressure
 (liquid-vapour) type phase transition and since in a star, the
pressure increases monotonically with density as we go from the 
surface to the core, mixed phase was strictly prohibited. It was 
pointed out by Glendenning \cite{3} that matter in neutron star has two components
, namely the conserved baryon number and the electric charge, therefore
the correct application of Gibb's phase rule is that the chemical
potential corresponding to baryon number and charge conservation
ie. $\mu_B$ and $\mu_Q$ , the temperature and the pressure in two 
phases are equal {\sl i.e.,}
$$\mu_B(h) = \mu_B(q) \quad \quad ; \quad \quad \mu_Q(h) = \mu_Q(q)$$
$$ p_h(\mu_B,\mu_Q,T) = p_q(\mu_B,\mu_Q,T)$$
and charge neutrality only demands Global conservation
$$ \chi Q_q(\mu_B,\mu_Q,T) + (1-\chi)Q_h(\mu_B,\mu_Q,T) =0$$
where $\chi$ is the fraction of the volume occupied by the quark phase. 
The freedom available to the system to rearrange concentration of charges
 for a given fraction of phases $\chi$, results in variation of the pressure
 through the mixed phase. 

\section{Radial Oscillations and Mass Radius}
The equations governing the radial oscillations of a non-rotating star, using 
static, spherically symmetric metric
\begin{equation}
d{s^2}=-e^{2\nu}dt^{2}+e^{2\lambda}dr^{2}+r^{2}(d{\theta}^{2}+sin^{2}{\theta}d\phi^{2})
\end{equation}
 were given by Chandrasekhar \cite{4}. The 
structure of the star in hydrostatic equilibrium is determined by the 
Tolman-Openheimer-Volkoff equations
\begin{equation}
\frac{dp}{dr}=\frac{-G(p+\rho)(m+4\pi r^3p)}{r^2(1-\frac{2GM}{r})}
\end{equation}
\begin{equation}
\frac{dm}{dr}=4\pi{r^{2}}\rho
\end{equation}
\begin{equation}
\frac{d\nu}{dr}=\frac{2GM(1+\frac{4\pi{r^3}p}{m})}{r(1-\frac{2GM}{r})}
\end{equation}
where we have put $ c = 1$.
Assuming a radial displacement ${\delta}r$ with harmonic time dependence 
${\delta}r$$\sim$$e^{i{\omega}t}$ and defining 
variables $\xi=\frac{{\delta}r}{r}$ and $\zeta=r^{2}e^{-\nu}\xi$ , 
the equation governing radial adiabatic oscillations is given by
\begin{equation}
F\frac{d^{2}\zeta}{dr^{2}}+G\frac{d\zeta}{dr}+H\zeta=\omega^{2}\zeta
\end{equation}
where
\begin{equation}
F=-\frac{e^{{2\nu}-{2\lambda}}({\gamma}p)}{p+\rho}
\end{equation}
\begin{equation}
G=-\frac{e^{{2\nu}-{2\lambda}}}{p+\rho}\Bigg[{\gamma}p(\lambda+3\nu)+\frac{d({\gamma}p)}{dr}-\frac{2}{r}({\gamma}p)\Bigg]
\end{equation}
\begin{equation}
H=\frac{e^{{2\nu}-{2\lambda}}}{p+\rho}\Bigg[\frac{4}{r}\frac{dp}{dr}+8{\pi}Ge^{2\lambda}p(p+\rho)-\frac{1}{p+\rho}(\frac{dp}{dr})^{2}\Bigg]
\end{equation}
$\lambda$ is related to the metric function through
\begin{equation}
e^{-2\lambda}=(1-\frac{2GM(r)}{r})
\end{equation}
and $\gamma$ is the adiabatic index, related to the speed of sound through
\begin{equation}
\gamma=\frac{p+\rho}{p}\frac{dp}{d\rho}
\end{equation}
Equation(5) is solved under the boundary conditions
\begin{equation}
\zeta(r=0) = 0 \hskip 1.5cm
{\delta}p(r=R) = 0
\end{equation}
where $\delta{p}(r)$ is given by
\begin{equation}
{\delta}p(r)=-\frac{dp}{dr}\frac{e^{\nu}\zeta}{r^{2}} - 
\frac{\gamma p e^{\nu}}{r^2}\frac{d\zeta}{dr}
\end{equation}
Equation (5) with the boundary condition (11) represent a Sturm-Liouville 
eigenvalue problem for $\omega^{2}$ with the well known result that the 
frequency 
spectrum is descrete. For $\omega^2 > 0$, $\omega$ is real and the solution is purely 
oscillatory
whereas for $\omega^2 < 0$, $\omega$ is imaginery resulting in exponentially growing unstable radial
oscillations. Another important consequence is that if the fundamental radial
mode $\omega_0$ is stable, so are the rest of the radial modes. For neutron stars
$\omega_0$ becomes 
imaginery at central densities $\rho_c > \rho_c^{critical}$ for which the star attains it's
maximum mass. For  $\rho_c = \rho_c^{critical}$, the fundamental frequency $\omega_0$ vanishes and becomes unstable for higher densities and the star is no longer stable.  There also
exists another unstable point at the lower end of the central density, namely
there exists a minimum mass for a stable neutron star and the frequency of
the fundamental mode at the minimum mass again goes to zero.

The only information required to obtain the structure of the star and 
eigenvalues of radial oscillation modes is the knowledge of the EOS. For a
given EOS, equations (2)-(5) are solved numerically by standard techniques
under the boundary conditions (11). While numerically integrating the 
equations for each EOS we make sure that the eigenfrequency of the fundamental
mode goes to zero at the maximum mass of the star. We have also checked that
the frequency  vanishes at the minimum stable mass too. For nuclear matter EOS
 we use two relativistic mean field theoretic
models taken from Glendenning \cite{5} and two potential models
incorporating relativistic corrections and three body interactions given by 
Akaml, Pandharipande and Ravenhall \cite{6}. Both class of models admit 
of mixed quark-nucleon phase in the core.
The quark matter in the above EOS models is described by the MIT Bag Model with 
$m_ u = m_ d =0$, $m_ s =150$ MeV, the Bag constant $B^{\frac{1}{4}} = 180$ 
MeV and $\alpha_s = 0$.
\begin{figure}[h]
\centerline{
\epsfig{file=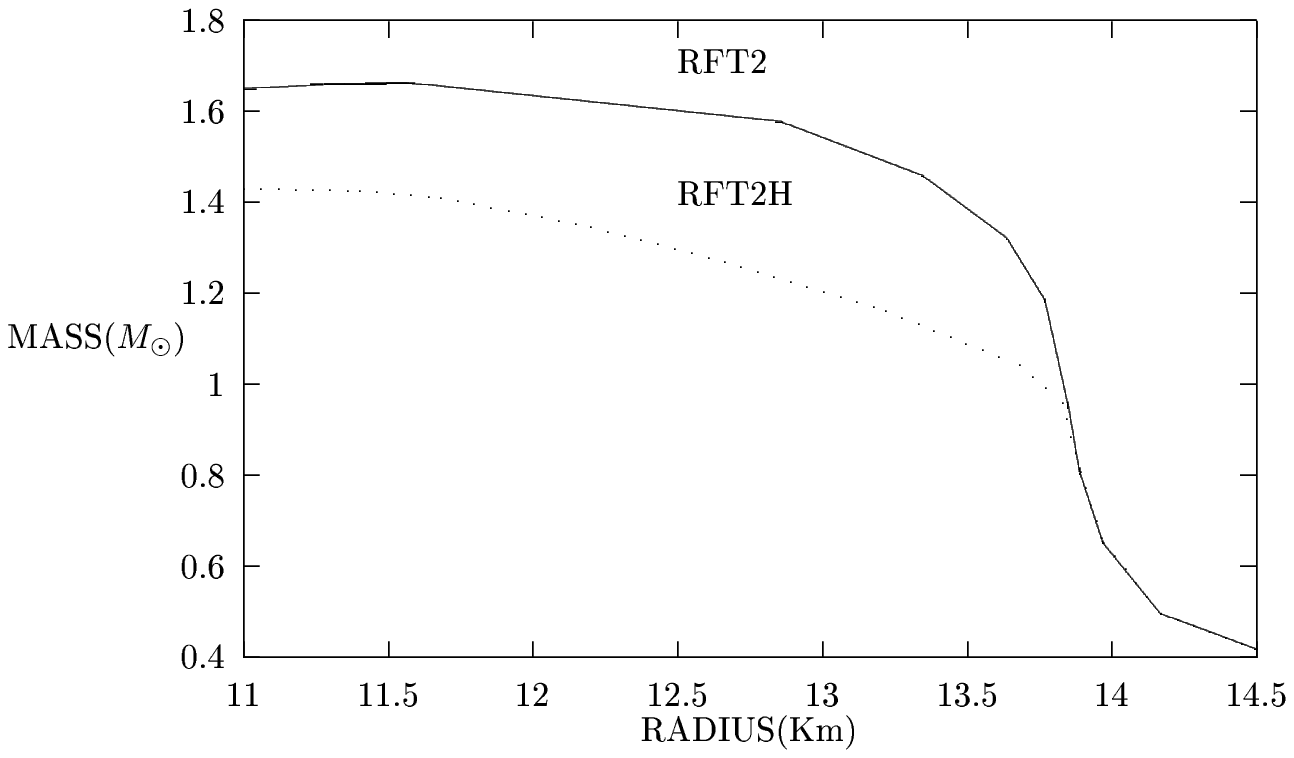,width=.48\textwidth}
\epsfig{file=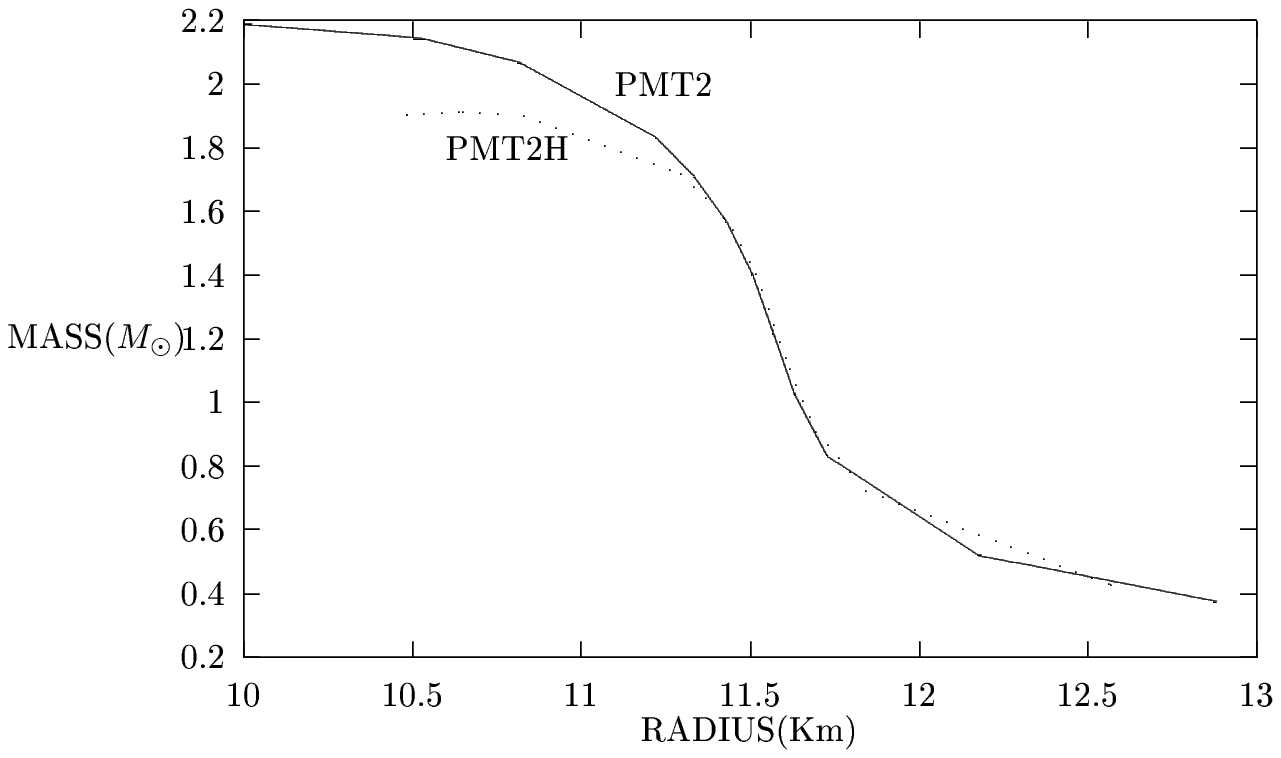,width=.48\textwidth}
}
\caption{Plot of mass in solar mass units versus radius in Km. Graphs
labbeled as in reference 7. }
\end{figure}
To illustrate the effect of mixed phase on neutron star parameters, namely 
mass-radius relationship and on the frequency of radial oscillations, in Fig.1
 we have plotted \cite{7} the M-R curves and find that the 
effect of the existence of 
mixed quark-nuclear matter phase in the core of neutron stars is to reduce the 
maximum mass. The effect is more pronounced for the relativistic mean field 
theory model than for the potential model. In Fig.2 we have plotted the 
frequency of the fundamental mode 
and the next mode as a function of central density for the pure neutron and 
hybrid 
stars in the two categories and find that the frequency exhibits oscillatory
behaviour in the case of a neutron star with a mixed quark-nuclear matter core.

The mass of a strange star in contrast to neutron star on the other hand 
 varies
as $\sim R^3$ for $M \sim 0.5 M_0$ and gravity plays minimal role, bag
pressure provides the confinement and the star is self bound. As mass
increases gravity becomes important and the star reaches a maximum
mass. In contrast to  neutron stars whose  radii increase with decreasing 
mass and
there is a minimum mass, strange star radii decreas with decrease in
mass and there is no minimum mass. Strange star density falls abruptly
to $\rho = 4B$ at the surface whereas in the the neutron star case
it falls to zero at the surface. The pure quark stars as expected are more compact
and they have much smaller radii with slightly smaller value of maximum mass
for the casewhen strange matter is modeled by the EOS given by Dey et al.
and for the CFL and superconducting phase. In the MIT bag model, there is little
difference in M-R of a  1.4
$M_0$ neutron or strange star. The effect of CFL/superconducting phase
in hybrid stars is similar to that shown in Figs.1 \& 2.
 The radial oscillation modes too exibit 
behaviour similar to that of neutron stars with similar values for the frequency \cite{8}.

\noindent Thus to summarize 
\begin{itemize}
\item{} The potential models give larger maximum stable mass upto 2.2 $M_0$ and higher
values of radial frequencies in contrast to relativistic mean field
models.  
\item{} The effect of the mixed phase is to soften the EOS thereby
lowering the maximum mass as well as the radial frequencies to the
extent of $\sim 30 \% $. 
\item{} For a 1.4 $M_0$ star, there is a substantial change in radial
frequency for the RFT models and none for PMT models. 
\item{} For a strange star, the radius decreases with mass and can be substantially
smaller in comparison to a neutron star radius and the radial
frequencies are of the same order as for neutron stars. 
\end{itemize}

\section{Observations and Conclusions }
During the last thirty years since the discovery of first pulsar in
1968 by Hewish et.al. close to two thousand pulsars using radio
telescope and X-ray probes in space have been discovered in a variety
of circumstances; 
\begin{itemize}
\item{} as isolated radio sources at times in binary
orbits with other stars
\item{} as X-ray pulsars and X-ray bursters in
X-ray binary system 
\item{} most recently by Rossi XTE X-ray
satellite as KHz quasi periodic oscillations (QPO) and burst oscillations in
LMXB's.
\end{itemize}
 A careful determination of the mass-radius 
relationship of these objects  has led not only to their identification
with neutron stars but has provided physical laboratory with
unprecedented potentialities to perform tests of GTR and to obtain
information on the EOS of high density matter its composition and
phase transition.The most accurate determination of neutron star masses is 
found in binary pulsars and the masses of all these neutron stars lie in the range
 $1.35 +0.04 M_0$ \cite{9}  
with exceptions of PSR J1012 of
mass $2.1 + 0.4 M_0$. Masses of X-ray pulsars are measured less
accurately and recent observations for Vela X-1 and Cygnus X-2 give
$1.9 + 0.2 M_0$ and $1.8 + 0.4 M_0$ respectively \cite{10}.
The recent discoveries of kilohertz-quasi-periodic
 oscillations in LMXB's provide a new method for determining masses and radii 
of neutron stars from the detection of X-ray pulsations by requiring that the
 inner radius of accretion flow $R_0$ be less than the radius of the star R 
but less than the corotation radius $R_c$ so that accretion is not 
centrifugally inhibited ie. $R<R_0<R_c$. Based on these considerations a 
mass-radius relationship  $R<8.54 (\frac{M}{M_0})^{1/2}$Km for SAX 
J1808.4-3658 has been obtained \cite{11}. Stringent constraints on the M-R 
relation \cite{12} have also been obtained for the X-ray sources 4U 1728-34
 ($M < 1.0 M_0$ and $R < 9$Km), for isolated compact star RX J1856-37
(M=0.9$\pm 0.2 M_0$,  
 R=6+2-3 Km), for X-ray pulsar Her X-1 (M=1.1-1.8 $M_o$, R=6-7.7 Km).
Most recently for the case of isolated neutron star RXJ1856 a radius $<6$Km has been 
obtained from the Chandra and XMM measurments of X-ray spectra \cite{13}.
\begin{figure}[h]
\centerline{
\epsfig{file=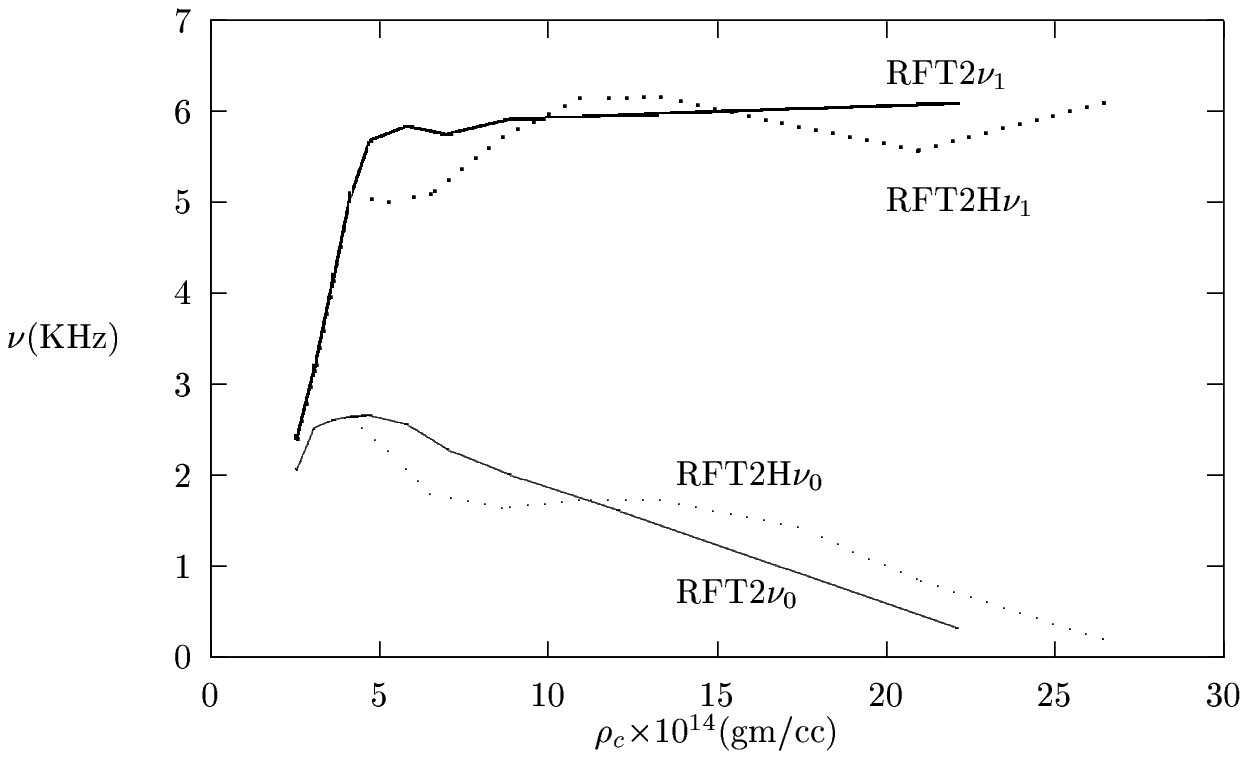,width=.48\textwidth}
\epsfig{file=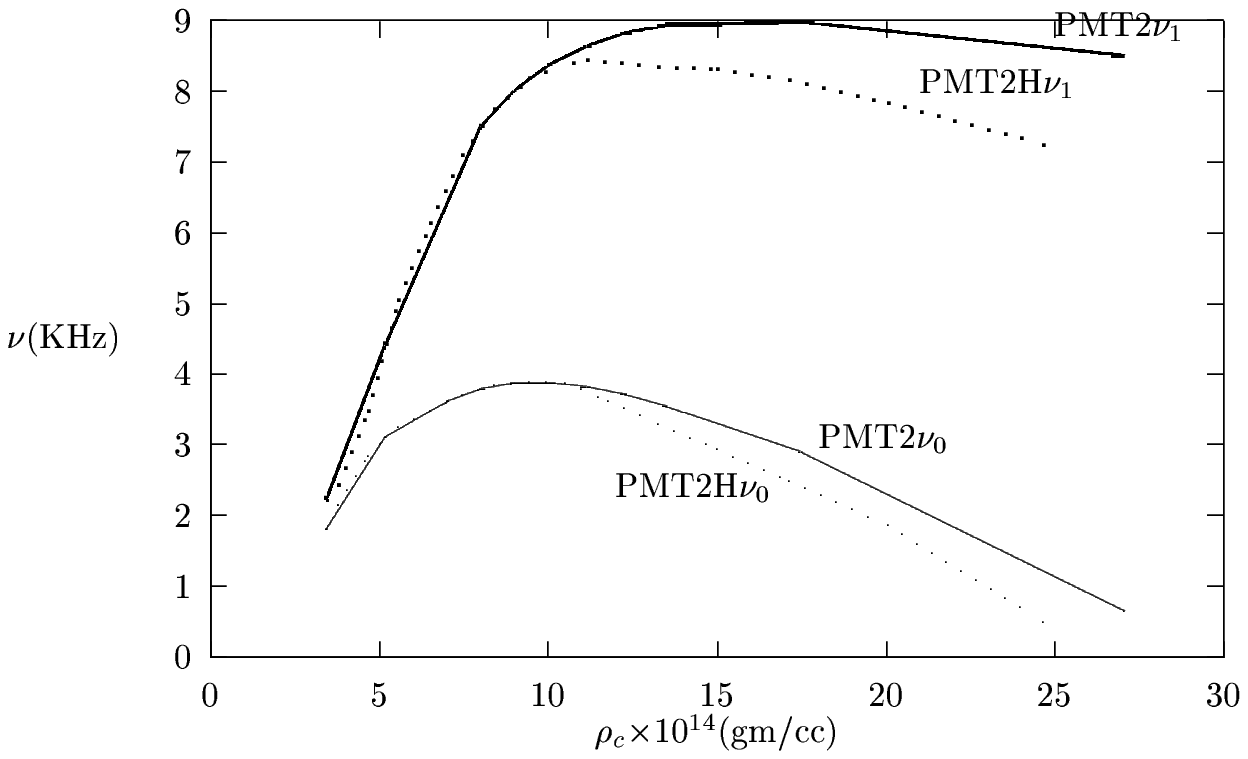,width=.48\textwidth}
}
\caption{Plot of frequency in KHz Vs central density in gm/cc}
\end{figure}
Thus
\begin{itemize}
\item{} If large masses of neutron stars are confirmed and complemented by
other neutron star masses $\sim 2 M_0$, EOS is severely restricted and
only stiff EOS's without any significant phase transition below $5n_0$
are allowed. In this scenario existence of strange stars is not required.
\item{} On the other hand if heavy neutron stars prove erroneous
by more detailed observations and masses like those of binary pulsars
($\sim 1.4 M_0$) alone are found , this will indicate that accretion does not produce
heavier stars which will mean either a soft EOS or significant phase
transition at few times the nuclear saturation density.
\item{} Observations of stars of mass $\leq M_0$ and radius $\leq 10
Km$ as seems to be borne out by  
present analysis of X-ray sources albeit with uncertainities in our knowledge 
of accretion mechanism and realistic EOS for the quark matter would imply 
the possibility of their identification with strange stars.
\end{itemize}

\section*{Acknowledgements}
I would like to thank the organisers of IWARA 2003 for inviting me to give 
this guest contribution. Partial support from DST and UGC, Delhi is acknowledged.


\begin{thebibliography}{0}
\bibitem{1} See for example K. Rajagopal and F. Wilczek, hep-ph/0011333 ; H. G. Alford, Ann. Rev. Nucl. Part. Sci. {\bf 51}, 131 (2001); H. G. Alford, hep-ph/0209287.
\bibitem{2} K. Schertler, C. Greiner and M. N. Thoma, Nucl. Phys.{\bf A616}, 659 (1997) ; M. Dey, I. Bombaci, J. Dey, S. Ray and B. C. Samanta, Phys. Lett. {\bf B438}, 123 (1998).
\bibitem{3} N. K. Glendenning, Phys. Rev. {\bf D46}, 1274 (1992).
\bibitem{4} S. Chandrasekhar, Phys. Rev. Lett., {\bf 140}, 417 (1964).
\bibitem{5} N. K, Glendenning, Compact stars, Springe (1967).
\bibitem{6} A. Akaml, V. J. Pandharipande and D. G. Ravenhall, Phys. Rev. 
{\bf 58}, 1804, (1998).
\bibitem{7} V. K. Gupta, Vinita Tuli and Ashok Goyal, Astrophys. J. {\bf 579}, 374 (2002).
\bibitem{8} B. Dutta, P. K. Sahu, J. D. Anand and Ashok Goyal, Phys. Lett. {\bf B283}, 313 (1992).
\bibitem{9} S. E. Thorsett and D. Chakraborty, Astrophys. J.,{\bf 512}, 288 (1999).
\bibitem{10} M. H.Van Kerkwijk,J. Van Paradijs and E. J. Zuiderwijk, A \& A, {\bf 303}, 497 
(1995) ; J. A. Orsoz, and E. Kuulkers, Mon. Not. R. Astron. Soc., {\bf 305}, 132 (1999).
\bibitem{11} X. D. Li, I. Bombaci, M. Dey, J. Dey and E. P. J. Van den
  Heuvel, Phys. Rev. Lett. {\bf 83}, 3776 (1999).
\bibitem{12} X. D. Li, S. Ray, J. Dey, M. Dey and I Bombaci,
  Astrophys. J. {\bf 527}, L51 (1999) ; J. A. Pons et al.,
  astro-ph/0107404 ; D. L. kaplan et al., astro-ph/0111174 ; S. M. Ransom et al., astro-ph/0111339.
\bibitem{13} J. J. Drake et al., Astrophys. J., {\bf 572}, 196 (2002).

\end{thebibliography}
\end{document}